\documentclass[12pt]{article}

\newcommand{\ie}{\textit{i.e.}}
\newcommand{\eg}{\textit{e.g.}}
\newcommand{\qinv}{q^{-1}}
\newcommand{\qana}{$q$-analog }
\newcommand{\sutwo}{$SU(2)$ }

\newcommand{\xth}{\frac{\partial\hat{x}}{\partial\theta}}
\newcommand{\xph}{\frac{\partial\hat{x}}{\partial\phi}}

\begin{document}
\begin{flushright}
\textbf{hep-th/9907153}
\end{flushright}

\vspace*{-.6in}
\thispagestyle{empty}

\vspace{.5in}
{\Large
\begin{center}
\textbf{One-parameter family of selfdual solutions in classical Yang-Mills theory}
\end{center}}
\medskip
\begin{center}
Masaru Kamata\footnote{e-mail: nkamata@minato.kisarazu.ac.jp}\\
\emph{Kisarazu National College of Technology,\\
 Kisarazu, 292-0041, Japan}\\
\end{center}
\begin{center}
and
\end{center}
\begin{center}
Atsushi Nakamula\footnote{e-mail: nakamula@sci.kitasato-u.ac.jp}\\
\emph{Department of Physics, School of Science\\ Kitasato University,\\ Sagamihara, 228-8555, Japan}\\
\end{center}
\vspace{0.5cm}

\begin{center}
\textbf{Abstract}
\end{center}

\noindent
The ADHM construction, which yields (anti-)selfdual configurations in classical Yang-Mills theories, is applied to an infinite dimensional $l^2$ vector space, and as a consequence, a family of (anti-)selfdual configurations with a parameter $q$ is obtained for \sutwo Yang-Mills theory.
This $l^2$ formulation can be seen as a \qana of Nahm's monopole construction, so that the configuration approaches the BPS monopole at $q\to1$ limit.

\begin{flushleft}
\end{flushleft}
\vfill

\thispagestyle{empty}
\newpage

\section{Introduction}
The Nahm equation\cite{Nahm82}, whose solutions yield monopole configurations in Yang-Mills(YM)/Higgs system, has a discrete analog, the discrete Nahm equation.
In ref.\cite{BA}, Braam and Austin give the correspondence between monopoles on hyperbolic three-space $H^3$, the hyperbolic monopole\cite{Ati}, and the solutions to the discrete Nahm equation.
In this paper, we introduce another way of the discretization to the Nahm formalism, which gives the solutions to the selfdual equation on Euclidean four-space, not the Bogomol'nyi equation on three-space with or without curvature.

The Nahm formalism is based on the construction of instantons in $Sp(n)$ YM theories by Atiyah, Drinfeld, Hitchin, and Manin(ADHM)\cite{AHDM}.
In the ADHM construction, an instanton goes with a finite dimensional vector space whose dimensionality concerns with the topological index of the instanton, for a review see \cite{Corri}.
Almost two decades ago, Nahm introduced infinite dimensional vector spaces associated  with each specific monopole configuration for $SU(2)\simeq Sp(1)$ Bogomol'nyi equation\cite{Nahm82}.
The vector spaces are quaternionic $\mathcal{L}^2\otimes V$ where the $\mathcal{L}^2$ is defined on a finite range, \eg, $[0,2]$, and $V$ is a finite dimensional space\cite{Hit83}.
In the present paper, we bring in,  instead of the $\mathcal{L}^2$ in the Nahm formalism, an $l^2$ vector space which is defined on $I_q$, a set of infinite discrete points in a finite range with a cluster point, \eg, $I_q=\{\pm\frac{1}{2},\pm\frac{1}{2}q,\pm\frac{1}{2}q^2,\cdots\}$ where $q\in(0,1)$ is a free parameter.
The operators acting on the $l^2$ space turn out to be $q$-derivative operators rather than differential ones for the case of $\mathcal{L}^2$.
We shall find that the ADHMN (ADHM and Nahm) construction  still works in our $q$-derivative case, the $l^2$ formulation, and that we can obtain a one-parameter family of (anti-)selfdual gauge fields.
Roughly speaking, the $l^2$ formulation is a \qana of ADHMN construction, so that the $q$-calculus  is requisite\cite{GR}.
This paper deals mainly with the explicit construction of (anti-)selfdual configuration, and we shall consider the physical interpretation and the mathematical detail of the configuration in a separate paper.

In the next section, we exhibit the one-parameter family of (anti-)selfdual configuration associated with the $l^2$ vector space.
In section 3, we summarize the $l^2$ formulation of ADHMN construction.
The last section is devoted to some discussion.

\section{The anti-selfdual configuration}
In line with the ADHMN construction, the gauge connections $A_\mu$ which make up (anti-)selfdual curvatures are given by a particular vector $v$ belonging to the appropriate vector space, such as
\begin{equation}
A_\mu=\,<v,i\partial_\mu v>, \label{gauge p}
\end{equation}
where the symbol $<\,\,\,,\,>$ denotes an inner product over the vector space under consideration.
In the next section, we perform the procedure to find the vector $v$ associated with $l^2$ vector space for the case of $SU(2)\simeq Sp(1)$ gauge group.
Here we exhibit the resulting gauge connection and curvature, with a parameter $q\in(0,1)$, which is the product of the \qana version of ADHMN construction.
The connection form is
\footnote{
Throughout the paper, we use the following conventions: $^\dagger$ denotes hermitian conjugation, $\tau_\mu=(1,i\sigma_1,i\sigma_2,i\sigma_3)$ and $x_\mu=(x_0,x_1,x_2,x_3)$ are quaternion elements and spacetime Euclidean coordinates, respectively, and $x:=\sum_{\mu=0}^3x_\mu\tau_\mu$.
We also use the three-space spherical coordinate $(r,\theta,\phi)$, and define $\hat{x}:=\sum_{j=1}^3x_j\sigma_j/r$.}
\begin{eqnarray}
A=\frac{1}{4}(-\frac{\partial\Omega}{\partial r}dx_0+\frac{\partial\Omega}{\partial x_0}dr)+\xi d\hat{x}-i\eta\hat{x}d\hat{x},\label{qbps}
\end{eqnarray}
where the matrix
\begin{equation}
\Omega(x_\mu;q):=\log\frac{L^+}{L^-}\cdot 1+\log(L^+L^-)\cdot\hat{x},\label{Omega}
\end{equation}
and the functions
\begin{eqnarray}
\xi(x_0,r;q)&:=&-\frac{i}{4}(M_+-M_-)(L^+L^-)^{-1/2},\\
\eta(x_0,r;q)&:=&-\frac{1}{2}\left\{1-\frac{M_++M_-}{2}(L^+L^-)^{-1/2}\right\}.
\end{eqnarray}
Here $L^\pm$ and $M_\pm$ are the functions of $\rho_\pm:=x_0\pm ir$ 
\begin{eqnarray}
L^\pm:=\sum^\infty_{n=0}\frac{(q\frac{\rho_\mp}{\rho_\pm};q)_{2n}}{(q^2;q)_{2n}}\left\{-\frac{(1-q)^2\rho_\pm^2}{4}\right\}^n, \label{Lpm}\\
M_\pm:=\sum^\infty_{n=0}\frac{1-q}{1-q^{2n+1}}\left\{-\frac{(1-q)^2\rho_\pm^2}{4}\right\}^n,\label{Mpm}
\end{eqnarray}
where the symbol $(a;q)_n$ is the $q$-shifted factorial,
\begin{equation}
 (a;q)_n:=\left\{ \begin{array}{lll} (1-a)(1-aq)(1-aq^2)\cdots(1-aq^{n-1})& 
\mbox{for} & n \ge 1, \\ 1 & \mbox{for} & n=0. \end{array}\right.
\end{equation}
One can straightforwardly derive the anti-selfdual curvature two-form\footnote{Although the connection (\ref{qbps}) is not $su(2)$-valued, we can obtain an $su(2)$-valued one through a ``gauge transformation" which preserves the curvature form.
We will discuss this point later.} from (\ref{qbps})
 and find their components
\begin{eqnarray}
F_{0r}&=&\left(\frac{\partial^2}{\partial x_0^2}+\frac{\partial^2}{\partial r^2}\right)\log(L^+L^-)\cdot\hat{x}\;,\label{F0r}\\
\nonumber\\
F_{\theta\phi}&=&-\frac{1}{2}\left(1-\frac{M_+M_-}{L^+L^-}\right)\sin\theta\cdot\hat{x}\;,\\
\nonumber\\
F_{0\theta}&=&\left\{\frac{\partial \xi}{\partial x_0}+\frac{1+2\eta}{4}\frac{\partial}{\partial r}\log(L^+L^-)\right\}\xth\nonumber\\
&&+\left\{\frac{\partial \eta}{\partial x_0}-\frac{\xi}{2}\frac{\partial}{\partial r}\log(L^+L^-)\right\}\frac{1}{\sin\theta}\xph\;,\\
\nonumber\\
F_{r\phi}&=&\left\{\frac{\partial \xi}{\partial r}-\frac{1+2\eta}{4}\frac{\partial}{\partial x_0}\log(L^+L^-)\right\}\xph\nonumber\\
&&-\left\{\frac{\partial \eta}{\partial r}+\frac{\xi}{2}\frac{\partial}{\partial x_0}\log(L^+L^-)\right\}\sin\theta\xth\;,\\
\nonumber\\
F_{r\theta}&=&\left\{\frac{\partial \xi}{\partial r}-\frac{1+2\eta}{4}\frac{\partial}{\partial x_0}\log(L^+L^-)\right\}\xth\nonumber\\
&&+\left\{\frac{\partial \eta}{\partial r}+\frac{\xi}{2}\frac{\partial}{\partial x_0}\log(L^+L^-)\right\}\frac{1}{\sin\theta}\xph\;,\\
\nonumber\\
F_{0\phi}&=&\left\{-\frac{\partial \eta}{\partial x_0}+\frac{\xi}{2}\frac{\partial}{\partial r}\log(L^+L^-)\right\}\sin\theta\xth\nonumber\\
&&+\left\{\frac{\partial \xi}{\partial x_0}+\frac{1+2\eta}{4}\frac{\partial}{\partial r}\log(L^+L^-)\right\}\xph\;,\label{F0p}
\end{eqnarray}
where the matrices $\hat{x},\; \xth,$ and $\frac{1}{\sin\theta}\xph$ span a representation of $su(2)$.
From these components we can explicitly confirm the anti-selfduality $F_{\mu\nu}=-\tilde F_{\mu\nu}$, and the hermiticity $F_{\mu\nu}^\dagger=F_{\mu\nu}$.

Although the gauge connection (\ref{qbps}) is in the form of infinite series, we have two configurations with elementary function expression by taking special values of the parameter $q$. 
One of them is the $q\to1$ limit, which gives the BPS monopole configuration\cite{Bogo,PS} if we regard the time component $A_0$ as an adjoint Higgs.
This can be seen from
\begin{equation}
L^\pm\to\frac{\sinh r}{r},\;M_\pm\to1.
\end{equation}
As we will see in the next section, this consequence is quite natural by construction.
The other is obtained if we take the $q\to0$ limit, because
\begin{equation}
L^\pm,\;M_\pm\to\left(1+\frac{\rho^2_\pm}{4}\right)^{-1}.
\end{equation}
The gauge configuration has infinite topological index also in this case.

\section{The $l^2$ formulation of ADHMN construction}
This section provides how to find out the vector $v$ in (\ref{gauge p}) by applying the $l^2$ formulation, concisely.

In $SU(2)$ YM theory, one can associate the instanton of topological index $k$ with a $(k+1)$-dimensional vector space.
For the monopole configurations, where we regard the $A_0$ component as an adjoint Higgs, Nahm introduced the infinite dimensional Hilbert space $\mathcal{L}^2$ instead of the finite dimensional vector space.
Since the ADHMN construction is established on quaternion field $H$, these vector spaces are all quaternionic.
The vector $v$ is to be determined through the linear operator $\Delta$ restricted to the conditions that $\Delta^{*}\Delta$ is quaternionic real and invertible, where $\Delta^*$ is an adjoint of $\Delta$. 
The ADHMN conditions on $v$ are
\begin{equation}
\Delta^{*}v=0, \label{Dv}
\end{equation}
and
\begin{equation}
<v,v>\,=1. \label{norm of v}
\end{equation}
In ref.\cite{Nahm80-1}  Nahm chose the operator $\Delta^{*}=i\partial_z+x^\dagger$ acting on  ${\cal L}^2[-1/2,1/2]\otimes H$ to obtain the BPS monopole configuration\cite{Bogo,PS}.

Now we shall consider another class of infinite dimensional vector space,  a vector space $l^2(I_q)\otimes H$ and define the operator $\Delta$ acting on the space.
As mentioned earlier, we specify the defining ``region" of $l^2(I_q)$ as $I_q:=\{\pm\frac{1}{2},\pm\frac{1}{2}q,\pm\frac{1}{2}q^2,\cdots\}$, $q$ being a real parameter of $0<q<1$, so that $l^2(I_q)\to\mathcal{L}^2[-1/2,1/2]$ at the limit $q\to1$.
Note that $I_q$ has a cluster point at zero.
We should emphasize that the discretization is multiplicative, which is distinct from the discrete Nahm case.
Next we define the inner product over the $l^2$ space.
Denoting a vector $v\in l^2[I_q]\otimes H$ as $v=f(x_\mu,z;q)$ where $z\in I_q$, let its adjoint $v^*$ be
\begin{equation}
v^*=[f(x_\mu,z;q)]^*\ =\ f^\dagger(x_\mu,qz;\qinv), \label{*}
\end{equation}
then we easily see $(v^{*})^{*}=v$.
Using this adjoint vector, we define our inner product
\begin{equation}
<w,v>\,=\int^{1/2}_{-1/2}w^* v \,d_qz:= \int^{1/2}_0w^* v \,d_qz-\int^{-1/2}_0w^* v \,d_qz, \label{<,>}
\end{equation}
where the integral is so called ``Thomae-Jackson integral" defined as
\begin{equation}
\int^a_{0}f(z) \,d_qz=a(1-q)\sum^\infty_{n=0}f(aq^n)q^n, \label{qint}
\end{equation}
so that (\ref{<,>}) is in fact the infinite summation over the discrete points of $I_q$.
Hereafter we call this ``integral" $q$-integral, for simplicity. 
Introducing the $q$-derivative operation
\begin{equation}
D_qf(z):=\frac{f(z)-f(qz)}{(1-q)z},\label{Dq}
\end{equation}
we can observe that our inner product makes the $q$-derivative operator $iD_q$  self-adjoint up to boundary terms, that is
\begin{equation}
<iD_qw,v>\,=\,<w,iD_qv>. \label{selfadj}
\end{equation}

Now we fix the linear operator $\Delta^*$ in (\ref{Dv}) as 
\begin{equation}
\Delta^*=iD_q+x^\dagger,\label{Dstar}
\end{equation}
then we have $\Delta=iD_q+x$ due to the self-adjointness of $iD_q$ under the $l^2$-inner product (\ref{<,>}).
The limit $q\to1$ leads to $\Delta^*\to i\partial_z+x^\dagger$, which brings the BPS monopole\cite{Nahm80-1}.
One can easily find the product $\Delta^*\Delta$ is quaternionic real,
\begin{eqnarray}
&\Delta^*\Delta=-D_q^2+2ix_0D_q+|x|^2&\nonumber\\
     &\ \qquad=(iD_q+\rho_{+})(iD_q+\rho_{-}),&
\end{eqnarray}
where $|x|^2:=\sum_{\mu=0}^3x_\mu^2$.
Thus if the invertible condition is fulfilled, we can obtain an (anti-)selfdual gauge configuration through the vector $v$ determined by (\ref{Dv}) and (\ref{norm of v}).
One can prove the invertibility by constructing the inverse function to $\Delta^*\Delta$ explicitly, \ie, finding out a function $F(x_\mu;z,z';q)$ with $z,z'\in I_q$ such that
\begin{equation}
\Delta^*\Delta F(x_\mu;z,z';q)=\delta_q(z,z').\label{DDinv}
\end{equation}
Here the right-hand side is a function taking non-zero value only when $z=z'$.
Since we are dealing with discrete space, the function $\delta_q(z,z')$ should be proportional to  Kronecker's delta $\delta_{z,z'}$.
Further, when we consider the limit $q\to1$, it should become the ordinary $\delta$-function in accordance with the BPS case.
We introduce a sign or step function defined as
\begin{equation}
\epsilon(z,z')=\left\{
	\begin{array}{ll}
	+1 & \mbox{if} \ z \ge z' \\
	-1 & \mbox{if} \ z < z' 
	\end{array}
\right. \; \mbox{for}\ 0<z',\; \mbox{and}\,
\left\{
	\begin{array}{ll}
	+1 & \mbox{if} \ z > z' \\
	-1 & \mbox{if} \ z \leq z' 
	\end{array}
\right. \; \mbox{for}\; z'<0.
\end{equation}
One can easily find that the $q$-derivative of $\epsilon(z,z')$ with respect to $z$ gives Kronecker's delta, \ie,
\begin{equation}
D_q\epsilon(z,z')=\frac{2}{(1-q)|z'|}\delta_{z,z'}.\label{kronecker}
\end{equation}
Designating the right hand side as the function $2\delta_q(z,z')$ in (\ref{DDinv}), and making use of the $q$-analogs of exponential function
\begin{eqnarray}
e_q(z)&=&\sum^\infty_{n=0}\frac{z^n}{(q;q)_n},\label{eq}\\
E_q(z)&=&\sum^\infty_{n=0}\frac{q^{n(n-1)/2}}{(q;q)_n}z^n,\label{Eq}
\end{eqnarray}
we obtain the solution to (\ref{DDinv}),
\begin{eqnarray}
&F(x_\mu;z,z';q)=\frac{1}{4r}\epsilon(z,z')\{E_q(-i\rho_+(1-q)qz')e_q(i\rho_+(1-q)z)\qquad&\nonumber\\
&\qquad -E_q(-i\rho_-(1-q)qz')e_q(i\rho_-(1-q)z)\}+F_0(x_\mu;z,z';q), &\label{FF}
\end{eqnarray}
where $F_0(x_\mu;z,z';q)$ is a kernel of $\Delta^*\Delta$ to be determined by boundary conditions, which does not need fixing here.
Thus we have proven the invertible condition.
Note that if we take the limit $q\to1$, the function $F$ comes exactly to be the one in Nahm's BPS derivation\cite{Nahm80-1}.
We therefore find that the inner product (\ref{<,>}) with (\ref{*}) is compatible with the ADHMN conditions.

Now, we solve (\ref{Dv}) and (\ref{norm of v}) to determine $v$.
Observing that one of the $q$-exponential functions (\ref{eq}) is the eigenfunction of $D_q$
\begin{equation}
D_qe_q(\alpha(1-q)z)=\alpha e_q(\alpha (1-q)z),
\end{equation}
where $\alpha$ is an arbitrary $z$-independent matrix, we find the functional form of $v$ as
\begin{equation}
v=e_q(ix^\dagger (1-q)z)N(x_\mu;q),\label{v=eN}
\end{equation}
$N(x_\mu;q)$ being a ``normalization function" to be determined by the condition (\ref{norm of v}).
Since both $e_q$ and $N$ are quaternion-valued functions, we have to take care the order of them, in general.
Next, let us fix the functional form of $N$.
First we observe the $l^2$ norm of $e_q(ix^\dagger (1-q)z)$ is
\begin{equation}
<e_q(ix^\dagger (1-q)z),e_q(ix^\dagger (1-q)z)>\,=\Lambda_+(x_0,r;q)\cdot1+\Lambda_-(x_0,r;q)\cdot\hat x,\label{e,e}
\end{equation}
where the functions $\Lambda_\pm$ are 
\begin{equation}
\Lambda_\pm(x_0,r;q)
=\frac{1-q}{2}\left\{\sum_{n=0}^\infty\frac{(\frac{\rho_+}{\rho_-};q)_{2n}}{(q;q)_{2n+1}}\biggl(i\frac{(1-q)\rho_-}{2}\biggr)^{2n}\pm (\rho_+\leftrightarrow\rho_-)\right\},\label{Apm}
\end{equation}
which are no longer quaternion-valued.
Note that $\hat x$ is an hermitian matrix depending only on the angles $\theta$ and $\phi$.
We roughly sketch  the derivation of (\ref{e,e}) and (\ref{Apm}) in the following.
The adjoint of $e_q$ is
\begin{equation}
[e_q(ix^\dagger(1-q)z)]^*=E_q(-ix(1-q)z).\label{estar=E}
\end{equation}
We can prove (\ref{estar=E}) by the fact
\begin{equation}
(q;q)_n=(\qinv;\qinv)_n(-q)^nq^{n(n-1)/2}.
\end{equation}
Thus if we directly apply the $q$-binomial theorem
\begin{equation}
\sum^\infty_{n=0}\frac{(a;q)_n}{(q;q)_n}z^n=\frac{(az;q)_\infty}{(z;q)_\infty},\label{qbin}
\end{equation}
where $(a;q)_\infty:=\prod_{m=0}^\infty(1-aq^m)$, to the product $E_q(-ix(1-q)z)e_q(ix^\dagger(1-q)z)$, we obtain the following formula after straightforward calculation
\begin{equation}
E_q(-ix(1-q)z)e_q(ix^\dagger(1-q)z)=\lambda_+(x_0,r;z;q)\cdot1+\lambda_-(x_0,r;z;q)\cdot\hat x,
\end{equation}
where
\begin{equation}
\lambda_\pm(x_0,r;z;q)=\frac{1}{2}\left\{\sum^\infty_{n=0}\frac{(\frac{\rho_+}{\rho_-};q)_n}{(q;q)_n}(i\rho_-(1-q)z)^n\pm(\rho_+\leftrightarrow\rho_-)\right\}.
\end{equation}
Here we used the facts
\begin{eqnarray}
x^m=\frac{1}{2}(\rho_+^m+\rho_-^m)\cdot1+\frac{1}{2}(\rho_+^m-\rho_-^m)\cdot\hat x,\\
x^{\dagger m}=\frac{1}{2}(\rho_+^m+\rho_-^m)\cdot1-\frac{1}{2}(\rho_+^m-\rho_-^m)\cdot\hat x,
\end{eqnarray}
for positive integer $m$.
Executing the $q$-integration,
\begin{equation}
\int^{1/2}_{-1/2}z^nd_qz=\left\{ \begin{array}{lll} 0& \mbox{for odd}& n, \\ 
\frac{1-q}{1-q^{n+1}}\frac{1}{2^n}& \mbox{for even}& n, \end{array}\right.
\end{equation}
for positive integer $n$, we obtain the result (\ref{e,e}) and (\ref{Apm}).
Next, we consider the normalization condition (\ref{norm of v}), which now turns out to be
\begin{equation}
N(x_\mu;q)^*(\Lambda_+\cdot1+\Lambda_-\cdot\hat x)N(x_\mu;q)=1.
\end{equation}
We can easily see that the inverse square root matrix $(\Lambda_+1+\Lambda_-\hat{x})^{-1/2}$ works well for $N$ due to the fact $\Lambda_\pm^*=\Lambda_\pm$, so that
\begin{equation}
N(x_\mu;q)=\frac{1}{2}\left[\{(L^+)^{-\frac{1}{2}}+(L^-)^{-\frac{1}{2}}\}\cdot 1+\{(L^+)^{-\frac{1}{2}}-(L^-)^{-\frac{1}{2}}\}\cdot\hat x\right], \label{N}
\end{equation}
where we defined $\Lambda_+\pm\Lambda_-=:L^\pm$ appeared in (\ref{Lpm}).
Note that we have $N^*=N$, trivially.

Having obtained the vector $v$, (\ref{v=eN}) with (\ref{N}), we can derive the gauge connection (\ref{qbps}) through (\ref{gauge p}) with the $q$-integral (\ref{<,>}).

\section{Discussion}
Our derivation fixes the functional form of $v$, and also the gauge connection $A_\mu$, up to ``gauge transformation" , which is realized by multiplying $v$  a matrix $g(x_\mu;q)$ on the right.
Under this operation, the gauge connection transforms
\begin{eqnarray}
A_\mu=\,<v,i\partial_\mu v> \rightarrow <vg,i\partial_\mu vg>\,=g^*A_\mu g + g^*i\partial_\mu g,\label{gauget}
\end{eqnarray}
thus  if $g(x_\mu,q)$ is subject to $g^*=g^{-1}$, then the curvature transforms covariantly and the YM action is invariant.
This implies our ``gauge transformation" (\ref{gauget}) is larger than the ordinary \sutwo.
As mentioned in section 2, the gauge connection (\ref{qbps}) is not $su(2)$-valued.
The part violating the $su(2)$-valuedness is the non-traceless and also non-hermitian term in (\ref{Omega}).
However, we can eliminate the non-$su(2)$ part without altering the $su(2)$-valued curvature form (\ref{F0r})--(\ref{F0p}) by performing a transformation with $g^*=g^{-1}$, \ie,
\begin{equation}
g(x_\mu;q)=\left[\left(-\frac{(1-q)^2}{4}\rho_+^2;q^2\right)_\infty\right]^{1/4}\left[\left(-\frac{(1-q)^2}{4}\rho_-^2;q^2\right)_\infty\right]^{1/4}\cdot1.\label{scale}
\end{equation}
This transformation has the form $g=e^{\chi\cdot1}$ with $\chi(\rho_\pm;q)$ being real function, which means (\ref{scale}) does not belong to $U(2)$.
Of course our curvature transforms covariantly under an ordinary, \ie, $q$-independent, $SU(2)$ transformation.

We can reduce one of the (anti-)selfdual conditions, $F_{0r}=-\tilde{F}_{0r}$, into the form of Liouville equation $\partial_+\partial_-\psi=2e^{-\psi}$, which is natural from the viewpoint of the earlier work on multi-instanton derivation\cite{W} and its equivalence to the ADHMN formalism.

To conclude, we have reformulated the ADHMN formalism on the $l^2$ vector space, and found the gauge field configuration, which reduces to the BPS monopole configuration at a limiting case of the parameter.
We can also formulate the $l^2$-analog of the Nahm equation, which will shed a new light on the consideration to spectral curves in monopole construction\cite{Hit83,HMM,HS96-1}, and to the integrability argument on the discrete Nahm equation\cite{Ward,MS}.

\end{document}